\documentclass[12pt,a4paper]{article}
\usepackage{amsmath}
\usepackage{amssymb}
\usepackage[cp866]{inputenc}
\usepackage[russian,ukrainian]{babel}

\begin{document}

\begin{center}
\textbf{On the Connection Between the Charged and Neutral Pion-Nucleon
Coupling Constants in the Yukawa Model}\bigskip

\textbf{V. A. Babenko\footnote{%
E-mail: pet2@ukr.net} and N. M. Petrov}

\bigskip

\textit{Bogolyubov Institute for Theoretical Physics, National Academy of
Sciences of Ukraine, Kiev}
\end{center}

\thispagestyle{empty}

\noindent In the Yukawa model for nuclear forces, a simple relation between
the charged and neutral pion-nucleon coupling constants is derived. The
relation implies that the charged pion-nucleon coupling constant is larger
than the neutral one since the $np$ interaction is stronger than the $pp$
interaction. The derived value of the charged pion-nucleon constant shows a
very good agreement with one of the recent experimental values. The relative
splitting between the charged and neutral pion-nucleon coupling constants is
predicted to be practically the same as that between the charged and neutral
pion masses. The charge dependence of the $NN$ scattering length arising
from the mass difference between the charged and neutral pions is also
analyzed.\bigskip

\bigskip

\begin{center}
1. INTRODUCTION\bigskip 
\end{center}

The pion-nucleon coupling constants are fundamental characteristics of
nuclear forces that play an important role in investigations of
nucleon-nucleon and pion-nucleon interactions [1--5]. Exact knowledge of
their values is instrumental for quntitatively describing and qualitatively
understanding a broad variety of hadron- and nuclear-physics phenomena
[3--8]. For this reason, the pion-nucleon coupling constants have been
investigated and their values have been refined throughout the whole period
of nuclear-physics studies. Historically, the evolution of these
investigations is detailed in [5, 7, 9].

That charge independence of the pion-nucleon coupling constant may be
violated, or pion-nucleon constants may differ for neutral and charged
pions, is a problem that has lately attracted much attention. That different
analyses yield different values of the charged pion-nucleon constant g$_{\pi
^{\pm }}^{2}$ renders this problem particularly important. Recent
experimental estimates of g$_{\pi ^{\pm }}^{2}$ vary between $13.54$ [9, 10]
and $14.74$ [11]. As for the neutral pion-nucleon constant g$_{\pi ^{0}}^{2}$%
, its value has been reliably and accurately measured as $13.5\div 13.6$\
[5, 9, 12--15].

Thus, the values of the charged-pion constant g$_{\pi ^{\pm }}^{2}$ measured
in [5, 7, 9, 10, 13, 16--19] are close to that of the neutral-pion constant g%
$_{\pi ^{0}}^{2}$ in agreement with the charge independence of the
pion-nucleon coupling. On the other hand, the measurements [5--7, 11, 14,
20--23] yielded significantly larger values of g$_{\pi ^{\pm }}^{2}$ than
that of g$_{\pi ^{0}}^{2}$. Therefore, the charge independence of the
pion-nucleon coupling is still an open fundamental problem which calls for
further experimental and theoretical investigation [5--7, 9--14, 18--30].

In this paper we investigate the pion-nucleon coupling constant relying on
the standard classical Yukawa model [1--3, 31] for the nucleon-nucleon
interaction and invoking contemporary experimental data on low-energy
parameters of nucleon-nucleon scattering. It is well known [1--3] that,
unfortunately, no accurate quantitative description of the nucleon-nucleon
system can be obtained with the Yukawa potential as soon as its parameters
are defined and selected using fundamental quantities of the field theory:
the pion masses and coupling constants. Therefore, in selecting the
parameter values of the Yukawa potential, we rather rely on measured
low-energy parameters of nucleon-nucleon scattering in the effective range
theory.

Analyzing the pion-nucleon coupling constants in the Yukawa-model framework
is justified, since the one-pion exchange which corresponds to the Yukawa
potential is the dominant mechanism of the nucleon-nucleon interaction at
the lowest collision energies, which imply long-range interactions. The
two-pion exchange and the heavier $\rho $- and $\omega $-meson exchanges are
dominant at medium and small distances, and quark-gluon degrees of freedom
gain an important role at the smallest interaction ranges. Quite
understandably, in determining the pion-nucleon coupling constant as a
characteristic of the pion-nucleon interaction, one should rely on the data
for long-range (or peripheral) nucleon-nucleon interaction, which is
dominated by one-pion exchange.

In quantum field theory, the pion-nucleon interaction may be described by
either a pseudoscalar ($PS$) or a pseudovector ($PV$) Lagrangian formulated
as [3, 9, 20, 32, 33]%
\begin{equation}
\mathcal{L} _{\pi N}^{PS}=\text{g}_{\pi }\sqrt{4\pi }\left( i\overline{\psi }%
\gamma _{5}\psi \right) \phi ~,  \tag{1}
\end{equation}%
\begin{equation}
\mathcal{L} _{\pi N}^{PV}=\frac{f_{\pi }}{m_{s}}\sqrt{4\pi }\left( \overline{%
\psi }\gamma _{\mu }\gamma _{5}\psi \right) \partial ^{\mu }\phi ~,  \tag{2}
\end{equation}%
where in the latter formula the scaling mass $m_{s}$ is introduced so as to
render the pseudovector pion-nucleon coupling constant $f_{\pi }$\
dimensionless. In Eqs. (1) and (2), the nucleon and pion fields are denoted
as $\psi \left( x\right) $ and $\phi \left( x\right) $, respectively. As was
demonstrated in [9, 34], the Lagrangian forms (1) and (2) are interrelated
by a gauge transformation and therefore are equivalent. Also note that our
Lagrangian definitions (1) and (2) feature explicit $\sqrt{4\pi }$ factors
so that the $4\pi $ factor no longer enters the coupling constants
themselves. In other words, we follow the convention adopted in the review
paper [9] and write simply g$_{\pi }^{2}$ instead of the often-used
denotation g$_{\pi }^{2}/4\pi $.

According to the convention adopted in [9, 32], the scaling mass $m_{s}$ is
usually assumed to be equal to the charged-pion mass: $m_{s}=m_{\pi ^{\pm }}$%
. Then, the pseudoscalar pion-nucleon coupling constant g$_{\pi }$ and the
pseudovector coupling constant $f_{\pi }$\ are interrelated by the
equivalence equation [9, 34] implied by the equivalence of Lagrangian forms
(1) and (2), 
\begin{equation}
\frac{\text{g}_{\pi }}{M_{1}+M_{2}}=\frac{f_{\pi }}{m_{\pi ^{\pm }}}~, 
\tag{3}
\end{equation}%
where $M_{1}$ and $M_{2}$ are the masses of interacting nucleons. Therefore,
the $\pi ^{0}$ and $\pi ^{\pm }$ pseudoscalar coupling constants, g$_{\pi
^{0}}$ and g$_{\pi ^{\pm }}$, and the corresponding pseudovector constants $%
f_{\pi ^{0}}$ and $f_{\pi ^{\pm }}$ may be interrelated through [5, 9, 14]%
\begin{equation}
\text{g}_{\pi ^{0}}=\frac{2M_{p}}{m_{\pi ^{\pm }}}~f_{\pi ^{0}}~,  \tag{4}
\end{equation}%
\begin{equation}
\text{g}_{\pi ^{\pm }}=\frac{M_{p}+M_{n}}{m_{\pi ^{\pm }}}~f_{\pi ^{\pm }}~,
\tag{5}
\end{equation}%
\ where $M_{p}$ and $M_{n}$ denote the proton and neutron masses,
respectively.\bigskip 

\begin{center}
2. DERIVATION AND DISCUSSION OF MAJOR EQUATIONS BETWEEN THE CHARGED AND
NEUTRAL PION-NUCLEON COUPLING CONSTANTS IN THE YUKAWA MODEL \bigskip 
\end{center}

According to the meson-field theory, strong interaction between two nucleons
at low energies is dominated by the exchange of virtual pions, which
determine the form of the long-range nucleon-nucleon interaction. The
classical one-pion-exchange potential of the nucleon-nucleon interaction in
the meson-field theory, referred to as the Yukawa potential, for the pure
singlet $^{1}S_{0}$ state has a simple and well-known form of [1--3, 31, 33]%
\begin{equation}
V_{Y}\left( r\right) =-V_{0}\frac{e^{-\mu r}}{\mu r}~.  \tag{6}
\end{equation}

Here, $r$ is the distance between the nucleons and $\mu $ is expressed
through the pion mass $m_{\pi }$ as%
\begin{equation}
\mu =\frac{m_{\pi }c}{\hbar }~,  \tag{7}
\end{equation}%
where $c$ and $\hbar $ are the speed of light and the reduced Planck
constant, respectively. According to (7), the nuclear-force range $R\equiv
1/\mu $ is inversely proportional to the pion mass $m_{\pi }$ and has a
small value of $R\thicksim 1.4~$fm. For this case, the depth of the Yukawa
potential, $V_{0}$\ , is expressed through the pseudovector pion-nucleon
coupling constant $f_{\pi }$ via a simple relation [1--3, 5, 33, 35]%
\begin{equation}
V_{0}=m_{\pi }c^{2}f_{\pi }^{2}~,  \tag{8}
\end{equation}%
which, like the Yukawa-potential form (6), is a direct consequence of the
quantum meson-field theory represented by the Lagrangians (1) and (2).

For the interaction between two charged protons mediated by the exchange of
a neutral $\pi ^{0}$ meson, the $\mu _{pp}$ and $V_{0}^{pp}$ parameters of
the Yukawa potential (6) are expressed through the $\pi ^{0}$ mass $m_{\pi
^{0}}$ and the coupling constant $f_{\pi ^{0}}$ according to equations (7)
and (8). The neutron-proton interaction involves the exchanges of both the
neutral $\pi ^{0}$ mesons and charged $\pi ^{\pm }$ mesons. In estimating
the $\mu _{np}$ and $V_{0}^{np}$ parameters of the potential (6) for the
latter case, one should [8, 36] substitute the averaged pion mass 
\begin{equation}
\overline{m}_{\pi }=\frac{1}{3}\left( m_{\pi ^{0}}+2m_{\pi ^{\pm }}\right) 
\tag{9}
\end{equation}%
and the averaged pion-nucleon coupling constant%
\begin{equation}
\overline{f_{\pi }^{2}}=\frac{1}{3}\left( f_{\pi ^{0}}^{2}+2f_{\pi ^{\pm
}}^{2}\right) ~.  \tag{10}
\end{equation}

We know [1--3] that, unfortunately, no accurate quantitative description of
the nucleon-nucleon system can be obtained with the Yukawa potential as soon
as its parameters are defined and selected using fundamental quantities of
the field theory --- the pion masses and coupling constants. Therefore, in
this paper the parameters of the Yukawa potential are assigned values
consistent with the measured parameters of low-energy nucleon-nucleon
scattering in the effective range theory [1, 4, 37--44].

Indeed, one may estimate the \textquotedblleft effective\textquotedblright\
mass $m_{\pi ^{0}}^{Y}$ and pion-nucleon coupling constant $\left( f_{\pi
^{0}}^{Y}\right) ^{2}$ for the neutral $\pi ^{0}$ meson from the measured
purely nuclear proton-proton scattering length and effective range, assuming
the Yukawa form for the proton-proton potential. Thus estimated $m_{\pi
^{0}}^{Y}$ and $\left( f_{\pi ^{0}}^{Y}\right) ^{2}$ values prove to
significantly exceed the directly measured values [26, 27], so that we have%
\begin{equation}
m_{\pi ^{0}}^{Y}=C_{1}m_{\pi ^{0}}\text{~},~~\left( f_{\pi ^{0}}^{Y}\right)
^{2}=C_{2}f_{\pi ^{0}}^{2}\text{~},  \tag{11}
\end{equation}%
where the factors $C_{1}$ and $C_{2}$ can be computed numerically [26, 27].
Their exact values are not needed for our purposes. It is quite natural to
assume that equations analogous to (11) also hold for the masses and
pion-nucleon coupling constants of charged $\pi ^{\pm }$ mesons, and
therefore also for the averaged pion mass (9) and averaged pion-nucleon
coupling constant (10).

Under the latter assumption and Eqs. similar in form to (11), from (8)--(10)
we obtain the following relations between the parameters of the
neutron-proton Yukawa potential, $\mu _{np}$ and $V_{0}^{np}$, and the
analogous parameters of the proton-proton potential, $\mu _{pp}$ and $%
V_{0}^{pp}$:%
\begin{equation}
\mu _{np}=\frac{\overline{m}_{\pi }}{m_{\pi ^{0}}}\mu _{pp}\text{~}, 
\tag{12}
\end{equation}%
\begin{equation}
V_{0}^{np}=\frac{\overline{m}_{\pi }}{m_{\pi ^{0}}}\frac{\overline{f_{\pi
}^{2}}}{f_{\pi ^{0}}^{2}}V_{0}^{pp}\text{~}.  \tag{13}
\end{equation}

Equations (10) and (13) directly imply a relation between the pseudovector
pion-nucleon coupling constants for the charged and neutral pions,%
\begin{equation}
f_{\pi ^{\pm }}^{2}=C_{f}^{2}f_{\pi ^{0}}^{2}\text{~},  \tag{14}
\end{equation}%
where the factor $C_{f}^{2}$ is expressed as%
\begin{equation}
C_{f}^{2}=\frac{1}{2}\left( 3\frac{m_{\pi ^{0}}}{\overline{m}_{\pi }}\frac{%
V_{0}^{np}}{V_{0}^{pp}}-1\right) \text{~}.  \tag{15}
\end{equation}

From Eqs. (4), (5), and (14) we obtain that the pseudoscalar charged and
neutral pion-nucleon coupling constants are interrelated as%
\begin{equation}
\text{g}_{\pi ^{\pm }}^{2}=C_{g}^{2}\text{g}_{\pi ^{0}}^{2}~,  \tag{16}
\end{equation}%
where%
\begin{equation}
C_{g}^{2}=\left( \frac{M_{p}+M_{n}}{2M_{p}}\right) ^{2}C_{f}^{2}~.  \tag{17}
\end{equation}

Equation (15) features directly measured pion masses, so in the considered
model the proportion between the charged and neutral pion-nucleon coupling
constants is fully determined by that between the depths of the
neutron-proton and proton-proton Yukawa potentials, $V_{0}^{np}/V_{0}^{pp}$.
Further we demonstrate that the neutron-proton potential is appreciably
deeper than the proton-proton one: $V_{0}^{np}>\frac{\overline{m}_{\pi }}{%
m_{\pi ^{0}}}V_{0}^{pp}$. As a consequence, the factor in parentheses in Eq.
(15) is in excess of two. Therefore, in the considered formalism, the
charged pion-nucleon coupling constant proves to be larger than the neutral
one:%
\begin{equation}
f_{\pi ^{\pm }}^{2}>f_{\pi ^{0}}^{2}\text{~},~~\text{g}_{\pi ^{\pm }}^{2}>%
\text{g}_{\pi ^{0}}^{2}\text{~}.  \tag{18}
\end{equation}

That the charged pion-nucleon constant is greater than the neutral one is,
within the considered scheme, a consequence of stronger $np$ than $pp$
interaction in the spin-singlet $^{1}S_{0}$ state, which is a reliably
established phenomenon. One of its manifestations is that the absolute
singlet length of $np$ scattering is larger than the purely nuclear $pp$
scattering length: $\left\vert a_{np}\right\vert >\left\vert
a_{pp}\right\vert $.

In a number of experiments [6, 11, 20--23], measured values of pion-nucleon
coupling constants obey the inequalities (18). On the other hand, the data
of other experiments [9, 10, 16--19] are consistent with charge independence
of the pion-nucleon constant, i.e., the (approximate) equalities $f_{\pi
^{\pm }}^{2}\cong f_{\pi ^{0}}^{2}$ and g$_{\pi ^{\pm }}^{2}\cong $g$_{\pi
^{0}}^{2}$ hold therein within the experimental uncertainties. In the
proposed model, charge dependence of nuclear forces reveals itself as a
violation of charge independence of the pion-nucleon coupling
constant.\bigskip 

\begin{center}
3. NUMERICAL RESULTS AND DISCUSSION OF CHARGE INDEPENDENCE BREAKING OF THE
PION-NUCLEON COUPLING CONSTANT\bigskip 
\end{center}

The depth $V_{0}^{NN}$ and radius $R_{NN}=1/\mu _{NN}$ of the Yukawa
nucleon-nucleon potential can be derived from the measured low-energy
parameters of the effective range expansion. Using the known values of
purely nuclear low-energy parameters of nucleon-nucleon scattering [4, 5,
35, 45--48]%
\begin{equation}
a_{pp}=-17.3(4)\,\text{fm~},~~r_{pp}=2.85(4)\,\text{fm,}  \tag{19}
\end{equation}%
\begin{equation}
a_{np}=-23.715(8)~\text{fm}  \tag{20}
\end{equation}%
and applying the variable phase approach [49], for the Yukawa-potential
parameters of $pp$ and $np$ interactions we obtain%
\begin{equation}
V_{0}^{pp}=44.8259\,\text{MeV~},~~\mu _{pp}=0.839241\,\text{fm}^{-1}\text{~},
\tag{21}
\end{equation}%
\begin{equation}
V_{0}^{np}=48.0706\,\text{MeV~},~~\mu _{np}=0.858282\,\text{fm}^{-1}\text{~}.
\tag{22}
\end{equation}

Note that the neutron-proton interaction parameters have been obtained using
Eq. (12) and substituting the neutron-proton scattering length as (20).

The Yukawa potential with parameters (22) results in an $np$-scattering
effective range of%
\begin{equation}
r_{np}=2.696\,\text{fm}  \tag{23}
\end{equation}%
which fully agrees with the measured value [41, 47, 50, 51]%
\begin{equation}
r_{np}=2.70(9)\,\text{fm~}.  \tag{24}
\end{equation}

As expected, we have%
\begin{equation}
V_{0}^{np}>\frac{\overline{m}_{\pi }}{m_{\pi ^{0}}}V_{0}^{pp}=45.8429\,\text{%
MeV~},  \tag{25}
\end{equation}%
so that the aforementioned condition leading to inequalities (18) is
satisfied.

Substituting the derived values (21) and (22) of the depths of $pp$ and $np$
potentials and the measured pion and nucleon masses [52] in (15) and (17),
for the factors relating the charged and neutral pion-nucleon coupling
constants we obtain%
\begin{equation}
C_{f}^{2}=1.0729\text{~},~~C_{\text{g}}^{2}=1.0744\text{~}.  \tag{26}
\end{equation}%
The $C_{f}^{2}$ and $C_{\text{g}}^{2}$ values are seen to be very similar,
as expected from the proximity of the proton and neutron masses.

In contrast with the charged pion-nucleon coupling constant g$_{\pi ^{\pm
}}^{2}$, the value of the neutral constant g$_{\pi ^{0}}^{2}$ has been
reliably measured and is not subject to controversy. One of the latest
measurements, g$_{\pi ^{0}}^{2}=13.52(23)$ [15], fully agrees with the
earlier experimental values of g$_{\pi ^{0}}^{2}=13.55(13)$ [12] and g$_{\pi
^{0}}^{2}=13.61(9)$ [13] and the mean value g$_{\pi ^{0}}^{2}=13.6(3)$
quoted in [5, 14]. Substituting in (16) the latest experimental value of the
pseudoscalar neutral constant%
\begin{equation}
\text{g}_{\pi ^{0}}^{2}=13.52(23)~~\text{[15]}  \tag{27}
\end{equation}%
and the $C_{\text{g}}^{2}$ value (26), for the pseudoscalar charged
pion-nucleon coupling constant we find%
\begin{equation}
\text{g}_{\pi ^{\pm }}^{2}=14.53(25)~.  \tag{28}
\end{equation}%
Using Eqs. (4), (5), (27), and (28), the pseudovector pion-nucleon coupling
constants are determined as%
\begin{equation}
f_{\pi ^{0}}^{2}=0.07479(127)~,  \tag{29}
\end{equation}%
\begin{equation}
f_{\pi ^{\pm }}^{2}=0.08027(138)~.  \tag{30}
\end{equation}

The g$_{\pi ^{\pm }}^{2}$ value (28) derived by us in the Yukawa-model
framework practically coincides with one of the most recent experimental
values:%
\begin{equation}
\text{g}_{\pi ^{\pm }}^{2}=14.52(26)~~\text{[6]}.  \tag{31}
\end{equation}%
The measurement (31), reported in [6] by the Uppsala Neutron Research Group,
is close to measurements of the same group, g$_{\pi ^{^{\pm
}}}^{2}=14.62(35) $ [23] and g$_{\pi ^{^{\pm }}}^{2}=14.74(33)$ [11], and to
the value g$_{\pi ^{^{\pm }}}^{2}=14.28(18)$ earlier obtained in [20--22].
On the other hand, the charged pion-nucleon coupling constant was extracted
by the Nijmegen group as g$_{\pi ^{^{\pm }}}^{2}=13.54(5)$ [9, 10], which
practically coincides with the neutral $\pi ^{0}$ constant g$_{\pi ^{0}}^{2}$%
. Other recent experimental evaluations [16--19] have yielded values of g$%
_{\pi ^{^{\pm }}}^{2}\sim 13.7\div 13.8$ which are close to the neutral-pion
constant g$_{\pi ^{0}}^{2}$. Thus, possible charge dependence of the
pion-nucleon coupling constant, or possible difference between those for
charged and neutral pions, is still an open problem which is of paramount
and fundamental importance. The proposed model involves an explicit
violation of charge independence of the pion-nucleon constant; see Eqs. (27)
and (28).

A measure of the charge independence breaking (CIB) of pion-nucleon
couplings is the difference between the charged and neutral pion-nucleon
coupling constants:%
\begin{equation}
\Delta f_{\text{CIB}}^{2}\equiv f_{\pi ^{\pm }}^{2}-f_{\pi ^{0}}^{2}\text{~}%
,~~\Delta \text{g}_{\text{CIB}}^{2}\equiv \text{g}_{\pi ^{\pm }}^{2}-\text{g}%
_{\pi ^{0}}^{2}\text{~}.  \tag{32}
\end{equation}%
In the considered model, Eqs. (14) and (16) imply the following explicit
expressions for these quantities:%
\begin{equation}
\Delta f_{\text{CIB}}^{2}=\left( C_{f}^{2}-1\right) f_{\pi ^{0}}^{2}\text{~},
\tag{33}
\end{equation}%
\begin{equation}
\Delta \text{g}_{\text{CIB}}^{2}=\left( C_{\text{g}}^{2}-1\right) \text{g}%
_{\pi ^{0}}^{2}\text{~}.  \tag{34}
\end{equation}%
Substituting the numerical value (26) for the factor $C_{\text{g}}^{2}$,
which relates the pseudoscalar charged and neutral pion-nucleon coupling
constants, as well as the reliably measured value (27) for the neutral
constant g$_{\pi ^{0}}^{2}$, for the absolute breaking of charge
independence of pion-nucleon coupling constants in the Yukawa model we obtain%
\begin{equation}
\Delta \text{g}_{\text{CIB}}^{2}=1.0055\text{~}.  \tag{35}
\end{equation}

In relative units, the charge independence breaking in pion-nucleon coupling
constants is formulated as%
\begin{equation}
\frac{\Delta f_{\text{CIB}}^{2}}{f_{\pi ^{0}}^{2}}=C_{f}^{2}-1=0.0729\text{~}%
,  \tag{36}
\end{equation}%
\begin{equation}
\frac{\Delta \text{g}_{\text{CIB}}^{2}}{\text{g}_{\pi ^{0}}^{2}}=C_{\text{g}%
}^{2}-1=0.0744\text{~}.  \tag{37}
\end{equation}

Thus, the relative violation of charge independence of pion-nucleon coupling
constants is as high as $7.4~\%$ in the considered scheme.

Equations (36) and (37) suggest that the relative charge independence
breaking is larger by $0.15~\%$ for the pseudoscalar pion-nucleon coupling
constant g$_{\pi }^{2}$ than for the pseudovector constant $f_{\pi }^{2}$.
According to Eqs. (15) and (17), this effect arises from the neutron-proton
mass difference ($M_{n}>M_{p}$). Therefore, even as soon as the pseudovector
coupling constant is strictly charge-independent ($f_{\pi ^{\pm
}}^{2}=f_{\pi ^{0}}^{2}$), charge invariance should be violated for the
pseudoscalar coupling constant g$_{\pi }^{2}$ [9].

It should be noted that the values (36) and (37) derived for relative charge
independence breaking of coupling constants are actually independent of
particular values of the charged and neutral pion-nucleon constants, but are
rather determined by the pion and nucleon masses according to (15) and (17)
and by the experimental input parameters of the model quoted in (19) and
(20).\bigskip 

\begin{center}
4. CONNECTION BETWEEN CHARGE SPLITTINGS OF THE PION-NUCLEON COUPLING
CONSTANT AND OF THE PION MASS\bigskip 
\end{center}

Equations (14) and (26) imply that the charged and neutral pseudovector
pion-nucleon coupling constants are in a ratio of%
\begin{equation}
\frac{f_{\pi ^{\pm }}}{f_{\pi ^{0}}}=C_{f}=1.0358\text{~},  \tag{38}
\end{equation}%
which closely agrees with the ratio between measured masses of the charged
and neutral pions [52]%
\begin{equation}
\frac{m_{\pi ^{\pm }}}{m_{\pi ^{0}}}=1.0340\text{~}.  \tag{39}
\end{equation}

Therefore, to a high precision we have%
\begin{equation}
\frac{f_{\pi ^{\pm }}}{f_{\pi ^{0}}}\cong \frac{m_{\pi ^{\pm }}}{m_{\pi ^{0}}%
}\text{~}.  \tag{40}
\end{equation}

Thus, charge splitting of the pion-nucleon coupling constant practically
coincides with that of the pion mass.

Equation (40) has a simple physical justification. Since the pion-nucleon
coupling constant $f_{\pi }$ is a measure of the strength of pion-field
action on the nucleon, this action is stronger the greater the pion mass $%
m_{\pi }$ is. Since we have $m_{\pi ^{\pm }}>$ $m_{\pi ^{0}}$, the nucleon
is more strongly affected by the ambient field of charged $\pi ^{\pm }$
mesons than by that of neutral $\pi ^{0}$ mesons. Equation (40) directly
implies that, to a high precision, the ratio $f_{\pi }/m_{\pi }$ is a charge
independent quantity, in contrast with the coupling constant $f_{\pi }$ by
itself.

As soon as the equality in (40) is exact, for the charged pseudovector
coupling constant one obtains%
\begin{equation}
f_{\pi ^{\pm }}^{2}=0.07997(136)\text{~}.  \tag{41}
\end{equation}

Under the same assumption, for the pseudoscalar coupling constant, Eqs. (5)
and (41) imply the value%
\begin{equation}
\text{g}_{\pi ^{\pm }}^{2}=14.48(25)~,  \tag{42}
\end{equation}%
which practically coincides with the value (28) computed with formula (16).
Then, from (27) and (42), the charge independence breaking in the
pion-nucleon coupling constant arising from the mass difference between the $%
\pi ^{\pm }$ and $\pi ^{0}$ mesons ($\Delta m_{\pi }=4.59~$MeV) is estimated
as%
\begin{equation}
\Delta \text{g}_{\text{CIB}}^{2}=0.96\text{~},  \tag{43}
\end{equation}%
which amounts to $\sim 7~\%$ in relative units.

Pseudoscalar coupling constants obey the approximate equation 
\begin{equation}
\frac{\text{g}_{\pi ^{\pm }}}{\text{g}_{\pi ^{0}}}\cong \frac{m_{\pi ^{\pm }}%
}{m_{\pi ^{0}}}\text{~},  \tag{44}
\end{equation}%
which is analogous to Eq. (40) for pseudovector coupling constants. Formula
(44) first attracted attention in [5, 14], but in these analyses it was
treated as an accidental coincidence which has no physical meaning. In our
analysis, Eqs. (40) and (44) result from implementing the traditional
classical Yukawa model and relying on measured values of low-energy $pp$-
and $np$-scattering parameters quoted in (19) and (20).

Equation (40) may be rewritten as%
\begin{equation}
f_{\pi ^{\pm }}R_{\pi ^{\pm }}\cong f_{\pi ^{0}}R_{\pi ^{0}}\text{~}, 
\tag{45}
\end{equation}%
where the radius $R_{\pi ^{0}}$ of the meson-exchange potential for the $\pi
^{0}$-exchange%
\begin{equation}
R_{\pi ^{0}}\equiv \frac{\hbar }{m_{\pi ^{0}}c}=1.4619\,\text{fm}  \tag{46}
\end{equation}%
is larger than the radius $R_{\pi ^{\pm }}$ corresponding to the $\pi ^{\pm
} $-exchange:%
\begin{equation}
R_{\pi ^{\pm }}\equiv \frac{\hbar }{m_{\pi ^{\pm }}c}=1.4138\,\text{fm~}. 
\tag{47}
\end{equation}

Thus, the pion-nucleon coupling constant $f_{\pi }$ and the radius of the
meson-exchange potential $R_{\pi }$ prove to be charge-dependent quantities
due to the mass splitting between the $\pi ^{\pm }$ and $\pi ^{0}$ mesons.
That $\pi ^{\pm }$ mesons are heavier than the $\pi ^{0}$ meson effectively
increases the charged-pion constant $f_{\pi ^{\pm }}$ with respect to the
neutral-pion constant $f_{\pi ^{0}}$ and effectively reduces the $\pi ^{\pm
} $-exchange radius $R_{\pi ^{\pm }}$ with respect to the $\pi ^{0}$%
-exchange radius $R_{\pi ^{0}}$. As a result, the product of the
pion-nucleon constant $f_{\pi }$ and the $\pi $-exchange radius $R_{\pi }$
is a charge-invariant quantity:%
\begin{equation}
f_{\pi }R_{\pi }=B\text{~}.  \tag{48}
\end{equation}

Substituting the well-measured value of the neutral pion-nucleon constant%
\begin{equation}
f_{\pi ^{0}}=0.2735  \tag{49}
\end{equation}%
and the value (46) for the $\pi ^{0}$-exchange radius, the constant $B$ is
numerically estimated as%
\begin{equation}
B=0.3998\,\text{fm~}.  \tag{50}
\end{equation}

Thus, the pion-nucleon coupling constant $f_{\pi }$ and the $\pi $-exchange
radius $R_{\pi }$ are correlated through the relation%
\begin{equation}
f_{\pi }\cong \frac{B}{R_{\pi }}~,  \tag{51}
\end{equation}%
which is valid to high precision.\bigskip 

\begin{center}
5. CHARGE DEPENDENCE OF THE NUCLEON-NUCLEON SCATTERING LENGTH\bigskip 
\end{center}

Since the $^{1}S_{0}$ state of the two-nucleon system features a virtual
level with a nearly zero energy, scattering length is the parameter which is
most sensitive to small variations of the nucleon-nucleon potential. For
this reason, the charge independence breaking of nuclear forces is often
quantitatively estimated using the difference between the proton-proton and
neutron-proton scattering lengths, 
\begin{equation}
\Delta a_{\text{CIB}}\equiv a_{pp}-a_{np}\text{~.}  \tag{52}
\end{equation}

According to (19) and (20), the experimental value of this difference is%
\begin{equation}
\Delta a_{\text{CIB}}^{\text{еxpt}}=6.42(41)~\text{fm~,}  \tag{53}
\end{equation}%
which amounts to $\sim 30~\%$ in relative units. That this difference is
nonzero significantly beyond the experimental uncertainty indicates that the
hypothesis of charge independence of nuclear forces is violated at low
energies [35, 53--56]. The charge dependence of nuclear forces is often
attributed to the mass difference between charged and neutral pions [14, 35,
53, 57--61]. However, only a half of the difference $\Delta a_{\text{CIB}}^{%
\text{еxpt}}$ has been shown to be due to the mass difference between the $%
\pi ^{\pm }$ and $\pi ^{0}$ mesons [14, 53, 60, 61].

The value of the singlet $np$-scattering length computed by us assuming that
equality (40) is exact, 
\begin{equation}
a_{np}=-22.89(40)~\text{fm~,}  \tag{54}
\end{equation}%
differs from the $pp$-scattering length of $a_{pp}=-17.3(4)~$fm. The
difference between the computed $pp$- and $np$-scattering lengths, 
\begin{equation}
\Delta a_{\text{CIB}}^{\pi }=5.59~\text{fm~,}  \tag{55}
\end{equation}%
is consistent with experimental value (53).

Thus, within the discussed model framework, the violation of charge
independence of nuclear forces is fully explained by the mass difference
between charged and neutral pions. The predicted difference between the $pp$%
- and $np$-scattering lengths, $\Delta a_{\text{CIB}}^{\pi }$, amounts to $%
\sim 90~\%$ of the corresponding experimental value $\Delta a_{\text{CIB}}^{%
\text{еxpt}}$. In contrast with this, the $\Delta a_{\text{CIB}}^{\pi }$
values derived in previous analyses reached only $\sim 50~\%$ of $\Delta a_{%
\text{CIB}}^{\text{еxpt}}$ [14, 60, 61].\bigskip 

\begin{center}
6. MAJOR CONCLUSIONS AND SUMMARY\bigskip 
\end{center}

In this paper, which is based on the Yukawa meson-field model, we develop a
physically consistent formalism of the nucleon-nucleon interaction in which
the parameters of the $np$ and $pp$ systems in the spin-singlet state $%
^{1}S_{0}$ are related to major characteristics of the pion-nucleon
interaction: the pion masses $m_{\pi }$ and the pion-nucleon coupling
constants $f_{\pi }^{2}$. Within this model framework, a simple relation
between the charged and neutral pion-nucleon coupling constants is
formulated by Eqs. (14)--(17).

The results of our analysis suggest that the charged pion-nucleon coupling
constant $f_{\pi ^{\pm }}^{2}$ is larger than the corresponding neutral
constant $f_{\pi ^{0}}^{2}$, so that charge independence of nuclear forces
is violated for the pseudovector and pseudoscalar pion-nucleon coupling
constants, $f_{\pi }^{2}$ and g$_{\pi }^{2}$.

Using the derived formulae, for the pseudoscalar charged constant we obtain
the value g$_{\pi ^{\pm }}^{2}=14.53(25)$, which virtually coincides with
the recent experimental value reported by the Uppsala Neutron Research Group
[6]: g$_{\pi ^{\pm }}^{2}=14.52(26)$.

The absolute value of the difference between the charged and neutral
pseudoscalar pion-nucleon coupling constants, $\Delta f_{\text{CIB}}\equiv $%
\ $f_{\pi ^{\pm }}-f_{\pi ^{0}}$, is derived by us as $\Delta f_{\text{CIB}%
}=0.0093$. In relative units, the ratio $\Delta f_{\text{CIB}}/f_{\pi
^{0}}=3.58~\%$ proves to be very close to $\Delta m_{\pi }/m_{\pi
^{0}}=3.40~\%$. Therefore, in relative units the charge splitting of the
pion-nucleon coupling constant is practically the same as that of the pion
mass.

Our analysis demonstrates that, while both the pion-nucleon coupling
constant $f_{\pi }$ and the $\pi $-meson-exchange radius $R_{\pi }$ are
charge dependent, their product $f_{\pi }R_{\pi }$ is, to high precision, a
charge independent quantity. In relative units, the difference between such
products for charged pions and neutral pions does not exceed $0.2~\%$.

In our model approach, $90~\%$ of the difference between experimental values
of the $pp$- and $np$-scattering lengths is accounted for by the mass
difference between the $\pi ^{\pm }$ and $\pi ^{0}$ mesons, $\Delta m_{\pi
}=4.59~$MeV.\bigskip

\begin{center}
{\normalsize REFERENCES\vspace{1pt}}
\end{center}

\begin{enumerate}
\item \textit{Hulth\'{e}n L., Sugawara M.} The Two-Nucleon Problem //
Encyclopedia of Physics, Vol. 39 -- Structure of Atomic Nuclei / Ed. S. Fl%
\"{u}gge. -- Berlin-G\"{o}ttingen-Heidelberg: Springer-Verlag, 1957. P.
1-143.

\item \textit{Bohr A., Mottelson B. R.} Nuclear Structure, Vol. 1. -- New
York: Benjamin, 1969. -- 471 P.

\item \textit{Ericson T., Weise W.} Pions and Nuclei. -- Oxford: Clarendon
Press, 1988. -- 479 P.

\item \textit{Miller G. A., Nefkens B. M. K., \v{S}laus I.} Charge Symmetry,
Quarks and Mesons // Phys. Rep. 1990. Vol. 194, No. 1-2. P. 1-116.

\item \textit{Machleidt R., Slaus I.} The Nucleon-Nucleon Interaction // J.
Phys. G. 2001. Vol. 27, No. 5. P. R69-R108.

\item \textit{Rahm J., Blomgren J., Cond\'{e} H., Dangtip S., Elmgren K.,
Olsson N., R\"{o}nnqvist T., Zorro R., Ringbom A., Tibell G., Jonsson O.,
Nilsson L., Renberg P.-U., Ericson T. E. O., Loiseau B.} np Scattering
Measurements at 162 MeV and the $\pi $NN Coupling Constant\ // Phys. Rev. C.
1998. Vol. 57, No. 3. P. 1077-1096.

\item \textit{Blomgren J., Ed. }Critical Issues in the Determination of the
Pion-Nucleon Coupling Constant: Proceedings of a Workshop Held in Uppsala,
Sweden, June 7-8, 1999. Phys. Scr. 2000. Vol. T87. P. 5-77. Uppsala: Royal
Swedish Academy of Sciences, 2000. 77 p.

\item \textit{Naghdi M.} Nucleon-Nucleon Interaction: A Typical/Concise
Review // Phys. Part. Nucl. 2014. Vol. 45, No. 5. P. 924-971.

\item \textit{de Swart J. J., Rentmeester M. C. M., Timmermans R. G. E.} The
Status of the Pion-Nucleon Coupling Constant // arXiv:9802084 [nucl-th],
1998. - 19 P.

\item \textit{Stoks V., Timmermans R., de Swart J. J.} Pion-Nucleon Coupling
Constant\ // Phys. Rev. C. 1993. Vol. 47, No. 2. P. 512-520.

\item \textit{Rahm J., Blomgren J., Cond\'{e} H., Dangtip S., Elmgren K.,
Olsson N., R\"{o}nnqvist T., Zorro R., Jonsson O., Nilsson L., Renberg
P.-U., Ringbom A., Tibell G., van der Werf S. Y., Ericson T. E. O., Loiseau
B.} np Scattering Measurements at 96 MeV // Phys. Rev. C. 2001. Vol. 63, No.
4. 044001.

\item \textit{Bergervoet J. R., van Campen P. C., Klomp R. A. M., de Kok
J.-L., Rijken T. A., Stoks V. G. J. , de Swart J. J.} Phase Shift Analysis
of All Proton-Proton Scattering Data Below $T_{\text{lab}}=350~$MeV\ //
Phys. Rev. C. 1990. Vol. 41, No. 4. P. 1435-1452.

\item \textit{Arndt R. A., Strakovsky I. I., Workman R. L.} Extraction of
the $\pi $NN Coupling Constant from NN Scattering Data\ // Phys. Rev. C.
1995. Vol. 52, No. 4. P. 2246-2249.

\item \textit{Machleidt R., Banerjee M. K.} Charge Dependence of the $\pi $%
NN Coupling Constant and Charge Dependence of the Nucleon-Nucleon
Interaction // Few-Body Syst. 2000. Vol. 28, No. 3. P. 139-146.

\item \textit{Limkaisang V., Harada K., Nagata J., Yoshino H., Yoshino Y.,
Shoji M., Matsuda M.} Phase-Shift Analysis of pp Scattering at $T_{L}=25-500$%
~MeV // Prog. Theor. Phys. 2001. Vol. 105, No. 2. P. 233-242.

\item \textit{Arndt R. A., Briscoe W. J., Strakovsky I. I., Workman R. L.,
Pavan M. M.} Dispersion Relation Constrained Partial Wave Analysis of $\pi $%
N Elastic and $\pi $N$\rightarrow \eta $N Scattering Data: The Baryon
Spectrum // Phys. Rev. C. 2004. Vol. 69, No. 3. 035213.

\item \textit{Arndt R. A., Briscoe W. J., Strakovsky I. I., Workman R. L.}
Extended Partial-Wave Analysis of $\pi $N Scattering Data // Phys. Rev. C.
2006. Vol. 74, No. 4. 045205.

\item \textit{Bugg D. V.} The Pion Nucleon Coupling Constant\ // Eur. Phys.
J. C. 2004. Vol. 33, No. 4. P. 505-509.

\item \textit{Baru V., Hanhart C., Hoferichter M., Kubis B., Nogga A.,
Phillips D. R.} Precision Calculation of Threshold $\pi ^{-}$d Scattering, $%
\pi $N Scattering Lengths, and the GMO Sum Rule // Nucl. Phys. A. 2011. Vol.
872, No. 1. P. 69-116.

\item \textit{Dumbrajs O., Koch R., Pilkuhn H., Oades G. C., Behrens H., de
Swart J. J., Kroll P.} Compilation of Coupling Constants and Low-Energy
Parameters\ // Nucl. Phys. B. 1983. Vol. 216, No. 2. P. 277-335.

\item \textit{Bugg D. V., Carter A. A., Carter J. R.} New Values of
Pion-Nucleon Scattering Lengths and $f^{2}$ // Phys. Lett. B. 1973. Vol. 44,
No. 3. P. 278-280.

\item \textit{Koch R., Pietarinen T.} Low-Energy $\pi $N Partial Wave
Analysis // Nucl. Phys. A. 1980. Vol. 336, No. 3. P. 331-346.

\item \textit{Ericson T. E. O., Loiseau B., Nilsson J., Olsson N., Blomgren
J., Cond\'{e} H., Elmgren K., Jonsson O., Nilsson L., Renberg P.-U., Ringbom
A., R\"{o}nnqvist T., Tibell G., Zorro R. }$\pi $NN Coupling from High
Precision np Charge Exchange at 162 MeV // Phys. Rev. Lett. 1995. Vol. 75,
No. 6. P. 1046-1049.

\item \textit{Alarcon J. M., Martin Camalich J., Oller J. A.} Improved
Description of the Scattering Phenomenology at Low Energies in Covariant
Baryon Chiral Perturbation Theory // Ann. Phys. 2013. Vol. 336. P. 413-461.

\item \textit{Matsinos E., Rasche G.} Analysis of the Low-Energy Charge
Exchange Data // Int. J. Mod. Phys. 2013. Vol. A28, No. 12. 1350039.

\item \textit{Babenko V. A., Petrov N. M.} Charge Dependence of the
Pion-Nucleon Coupling Constant // Nuclear Physics and Atomic Energy. 2015.
Vol. 16, No. 2. P. 136-143. [In Russian]

\item \textit{Babenko V. A., Petrov N. M.} Study of the Charge Dependence of
the Pion-Nucleon Coupling Constant on the Basis of Data on Low-Energy
Nucleon-Nucleon Interaction // Phys. Atom. Nucl. 2016. Vol. 79, No. 1. P.
67-71; \textit{Babenko V. A., Petrov N. M.} Study of the Pion-Nucleon
Coupling Constant Charge Dependence on the Basis of the Low-Energy Data on
Nucleon-Nucleon Interaction // arXiv:1604.02912 [nucl-th], 2016. - 10 P.

\item \textit{Arriola E. R., Amaro J. E., Perez R. N.} Three Pion Nucleon
Coupling Constants // Mod. Phys. Lett. 2016. Vol. A31, No. 28. 1630027.

\item \textit{Matsinos E., Rasche G.} Systematic Effects in the Low-Energy
Behavior of the Current SAID Solution for the Pion-Nucleon System // Int. J.
Mod. Phys. 2017. Vol. E26, No. 3. 1750002.

\item \textit{Matsinos E., Rasche G.} Update of the Phase-Shift Analysis of
the Low-Energy Data // arXiv:1706.05524 [nucl-th], 2017. - 76 P.

\item \textit{Yukawa H.} On the Interaction of Elementary Particles // Proc.
Phys. Math. Soc. Jap. 1935. Vol. 17. P. 48-57.

\item \textit{Ebel G., M\"{u}llensiefen A., Pilkuhn H., Steiner F., Wegener
D., Gourdin M., Michael C., Petersen J. L., Roos M., Martin B. R., Oades G.,
De Swart J. J. }Compilation of Coupling Constants and Low-Energy Parameters\
// Nucl. Phys. B. 1971. Vol. 33, No. 2. P. 317-378.

\item \textit{Bjorken J. D., Drell S. D.} Relativistic Quantum Mechanics. --
New York: McGraw-Hill, 1964. -- 300 P.

\item \textit{Dyson F. J.} The Interactions of Nucleons with Meson Fields //
Phys. Rev. 1948. Vol. 73, No. 8. P. 929-930.

\item \textit{Sliv L. A.} Charge Independence and Charge Symmetry of Nuclear
Forces // Izv. Akad. Nauk SSSR, Ser. Fiz. 1974. Vol. 38, No. 1. -- P. 2 --
14. [In Russian]

\item \textit{Ericson T. E. O., Rosa-Clot M.} The Deuteron Asymptotic
D-state as a Probe of the Nucleon-Nucleon Force // Nucl. Phys. A. 1983. Vol.
405, No. 3. P. 497-533.

\item \textit{Landau L. D., Smorodinsky Ya. A.} Scattering of Protons by
Protons // Zh. Eksp. Teor. Fiz. 1944. Vol. 14, No. 7-8. P. 269-278. [In
Russian]

\item \textit{Schwinger J. S.} A Variational Principle for Scattering
Problems // Phys. Rev. 1947. Vol. 72, No. 8. P. 742.

\item \textit{Blatt J. M., Jackson J. D.} On the Interpretation of
Neutron-Proton Scattering Data by the Schwinger Variational Method // Phys.
Rev. 1949. Vol. 76, No. 1. P. 18-37.

\item \textit{Bethe H. A.} Theory of the Effective Range in Nuclear
Scattering // Phys. Rev. 1949. Vol. 76, No. 1. P. 38-50.

\item \textit{Sitenko A. G., Tartakovskii V. K.} Lectures on the Theory of
the Nucleus. -- Oxford, New York: Pergamon Press, 1975. -- 304 P.

\item \textit{Pupyshev V. V., Solovtsova O. P.} Long-Range Potentials in
Low-Energy Nuclear Physics // Phys. Part. Nucl. 1996. Vol. 27, No. 4. P. 353.

\item \textit{Pupyshev V. V.} Low-Energy Expansions in Nuclear Physics //
Phys. Part. Nucl. 1997. Vol. 28, No. 6. P. 586.

\item \textit{Babenko V. A., Petrov N. M.} Determination of Low-Energy
Parameters of Neutron-Proton Scattering on the Basis of Modern Experimental
Data from Partial-Wave Analyses // Phys. Atom. Nucl. 2007. Vol. 70, No. 4.
P. 669-675; \textit{Babenko V. A., Petrov N. M.} On Triplet Low-Energy
Parameters of Nucleon-Nucleon Scattering // Phys. Atom. Nucl. 2006. Vol. 69,
No. 9. P. 1552-1572.

\item \textit{Houk T. L., Wilson R.} Erratum: Measurements of the
Neutron-Proton and Neutron-Carbon Cross Sections at Electron Volt Energies
// Rev. Mod. Phys. 1968. Vol. 40, No. 3. P. 672.

\item \textit{Hackenburg R. W.} Neutron-Proton Effective Range Parameters
and Zero-Energy Shape Dependence // Phys. Rev. C. 2006. Vol. 73, No. 4.
044002.

\item \textit{Babenko V. A., Petrov N. M.} Determination of Low-Energy
Parameters of Neutron-Proton Scattering in the the Shape-Parameter
Approximation from Present-Day Experimental Data // Phys. Atom. Nucl. 2010.
Vol. 73, No. 9. P. 1499-1506.

\item \textit{Babenko V. A., Petrov N. M.} Mirror Nuclei $^{3}$H and $^{3}$%
He Binding Energies Difference and Low Energy Parameters of Neutron-Neutron
Scattering // Physics of Particles and Nuclei Letters. 2015. Vol. 12, No. 4.
P. 584-590.

\item \textit{Babikov V. V.} Variable Phase Approach in Quantum Mechanics.
-- Moskow: Nauka, 1976. -- 288 P. [In Russian]

\item \textit{Breit G., Friedman K. A., Holt J. M., Seamon R. E.} Short- and
Long-Range Charge Independence // Phys. Rev. 1968. Vol. 170, No. 5. P.
1424-1434.

\item \textit{Lock W. O., Measday D. F.} Intermediate Energy Nuclear
Physics. -- London: Methuen, 1970. -- 320 P.

\item \textit{Beringer J.} \textit{et al. (Particle Data Group).} Review of
Particle Physics // Phys. Rev. D. 2012. Vol. 86, No. 3. 010001.

\item \textit{Kuhn B.} Measurements of the Neutron-Neutron Scattering Wave
Length and the Problem of Charge Dependence of Nuclear Forces // Phys. Part.
Nucl. 1975. Vol. 6, No. 2. P. 347-392.

\item \textit{Aleksandrov Yu. A.} Fundamental Properties of the Neutron. --
Oxford, UK: Clarendon Press, 1992. -- 210 P.

\item \textit{Popov Yu. P., Sharapov E. I.} To the Question on the
Implementation of the Charge Invariance Principle in Nuclear Interactions //
Strong and Weak Statements in Nuclear Spectroscopy and Nuclear Theory.
Leningrad: Nauka, 1981. P. 90-95. [In Russian]

\item \textit{Miller G. A., van Oers W. T. H.} Charge Independence and
Charge Symmetry // arXiv:9409013 [nucl-th].

\item \textit{Sugie A.} The Effect of the Mass Difference Between Charged
and Neutral Pions on the Nuclear Force // Prog. Theor. Phys. 1954. Vol. 11,
No. 3. P. 333-334.

\item \textit{Riazuddin. }On the Charge Independence of Nuclear Forces //
Nucl. Phys. 1956/57. Vol. 2. P. 188-191.\textit{\ }

\item \textit{Riazuddin. }Charge Dependent Effects on Scattering Lengths of
np and pp Systems // Nucl. Phys. 1958. Vol. 7. P. 217-222.

\item \textit{Henley E. M., Morrison L. K.} n-n and n-p Scattering Lengths
and Charge Independence // Phys. Rev. 1966. Vol. 141, No. 4. P. 1489-1493.

\item \textit{Ericson T. E. O., Miller G. A.} Charge Dependence of Nuclear
Forces // Phys. Lett. B. 1983. Vol. 132, No. 1--3. P. 32-38.
\end{enumerate}

\end{document}